 \documentclass{aa}

\begin{document}

 \thesaurus{
 12.12.1,  
 10.05.1   
 }

 \title{
 Multicolour Observations, Inhomogeneity \& Evolution 
 }

 \author{
 Charles Hellaby
 \inst{1}
 }

 \institute{
 Department of Mathematics and Applied Mathematics, 
 University of Cape Town, Rondebosch, 7701, South Africa. 
 email: cwh@maths.uct.ac.za
             }

 \date{
 Submitted to A\&A: 31 October 2000 \\
 arXiv Ref:~ astro-ph/0010641 \\
 Report No:~ uct-cosmology-00/09
 }

 \maketitle

 \begin{abstract}
 We propose a method of testing source evolution theories that is 
independent of the effects of inhomogeneity, and thus complementary to 
other studies of evolution.  It is suitable for large scale sky surveys, 
and the new generation of large telescopes.  
 In an earlier paper it was shown that basic cosmological observations 
 --- luminosity versus redshift, area distance versus redshift and number 
counts versus redshift 
 --- cannot separate the effects of cosmic inhomogeneity, cosmic evolution 
and source evolution.  We here investigate multicolour observations, and 
show that by comparing luminosity versus redshift in two or more colours, 
contraints can be placed on source evolution even if unknown source 
evolution is present, providing an important test of evolution theories 
that is complementary to present methods.  However, number counts in 
different colours versus redshift are not useful in separating the effects 
of source evolution and inhomogeneity. 



 \end{abstract}

 \section{Introduction} 

 As measurements of the Cosmos become more extensive and more accurate, it 
becomes increasingly important to take account of the evolution of 
galaxies and clusters of galaxies, since these have a direct effect on our 
cosmological measurements.  It seems that the Hubble constant may soon be 
well known, and limits on the acceleration parameter may improve rapidly.  
The next step is to estimate deviations from homogeneity on the large 
scale%
 \footnote{ 
 For example, Wang, Spergel and Turner (\cite{WST98}) have estimated that 
variations of up to 10\% between local measures of $H_0$ and its large 
scale value are possible. 
 }%
 .  It is expected that the Sloan digital sky survey (SDSS) will detail 
the galaxy distribution out to $z \leq 0.5$.  How does this data relate to 
the spacetime metric? 

 It was previously shown by Mustapha Hellaby and Ellis (\cite{MHE97}) 
(MHE) that observations of luminosity versus redshift and area distance 
(or luminosity distance) versus redshift cannot distinguish between the 
effects of source evolution, cosmic inhomogeneity, and cosmic evolution.  
Riess et al (\cite{RFCea98}) and Perlmutter et al (\cite{PAGea99}) have 
fitted supernova observations to an FLRW model with non-zero $\Lambda$, 
but Celerier (\cite{C00}) has shown they could equally well be fitted by 
an inhomogeneous model with zero $\Lambda$%
 \footnote{
 Also Maor, Brustein and Steinhardt (\cite{MBS00}) argue these 
observations cannot determine the future fate of the universe. 
 }%
 .

 Whilst the Copernican Principle may lead many to assert that homogeneity 
is valid on a large enough scale, it is better not to assume something if 
it can be measured.  In addition, the observed luminosity versus redshift 
relation does not fit the FLRW predictions without some adjustments, such 
as a cosmological constant that has a problematic value, or source 
evolution.  And since source evolution is really not known, one should be 
suspicious of the results.  Thus it has never been observationally 
determined on which scale homogeneity is valid%
 \footnote{
 For example, Barrett \& Clarkson (\cite{BC99}) have shown that Stephani 
models with significant inhomogeneity can fit a range of key cosmological 
observations. 
 }%
 . 

 Galaxy evolution is now a very active field, but there is a long way to 
go%
 \footnote{
 see e.g. Ellis, Abraham, Brinchmann and Menanteau (\cite{EABM00}) and 
references therein, Kennicut (\cite{K98}), Ellis (\cite{E97}), Impey and 
Bothun (\cite{IB97}), Bell and Bower (\cite{BB00}), Blundell and Rawlings 
(\cite{BR99}), Brunner Szalay and Connolly (\cite{BSC00}), Bullock Dekel 
and Primack (\cite{BDP00}), Bunker et al (\cite{BSSTMDD00}), Cavaliere and 
Vittorini (\cite{CV00}), Dickinson (\cite{D00}), Fasano et al 
(\cite{FPCBKM00}), Genzel Lutz and Tacconi (\cite{GLT98}), Kodama and 
Bower (\cite{KB00}), Ponman Cannon and Navarro (\cite{PCN99}), Watts and 
Taylor (\cite{WT00}). 
 }%
 .  There are many functions to be determined 
 --- the rate of formation of each galaxy type, the rate of formation of 
stars of different masses in each galaxy type the rate of galaxy mergers%
 \footnote{ 
 It has also been proposed that stellar mergers may be significant in 
dense clusters. 
 }%
 , the effect of galaxy mergers and encounters on star formation, how 
central bulges and bars form, how low surface brightness galaxies fit into 
the picture, etc.  The early appearance of galaxies and the relationship 
between quasars and galaxies is really not known.  Many studies of source 
evolution make an assumption of homogeneity at some point, by using 
observational relations derived in a FLRW model.  While we don't dispute 
their usefulness, the danger is that the resulting source evolution 
relations may later be used to demonstrate homogeneity.  Thus methods of 
determining source evolution that don't make assumptions about the 
universe model should be emphasised as more reliable.  This point was also 
made by Goodman (\cite{G95}).  Observations of supernovae (Riess et al 
\cite{RFCea98}, Perlmutter et al \cite{PAGea99}) may help to separate 
inhomogeneity from source evolution, out to distances where their light 
curves can be measured, though it is still uncertain whether they are free 
of evolution effects (e.g. Riess et al \cite{RFLS99}). 

 As a contribution towards this requirement, we investigate whether 
multicolour observations can help to separate inhomogeneity from source 
evolution.  The measurements in different colours may be a selection of 
spectral line intensities or luminosties in U, B \& V filters, say.  In 
any case, one needs to know the absolute luminosity in each colour at each 
$z$ value, so there is a new source evolution function for each colour.  
Thus it seems that multicolour observations may not improve the situation.  
However, a key point is that the luminosity distance must be the same in 
all wavelengths for each given source.  So if the 
 luminosity-redshift plot in two colours is not the same, this is evidence 
that the two colours have different evolution functions.  This approach is 
used below, and it is found that the basic uncertainty in absolute values 
remains, but the 
 luminosity-redshift relations in different colours must be related to 
each other, thus providing constraints on evolution theories. 

 The choice of cosmological model is not central to the ideas presented 
here, but to keep the equations simple and focus on the basic concepts, we 
choose the simplest inhomogeneous cosmological solution of Einstein's 
equations.  This is the 
 Lema\^{\i}tre-Tolman (LT) model%
 \footnote{Though it is sometimes called the 
 Lema\^{\i}tre-Tolman-Bondi (LTB) model, Krasinski's (\cite{K97}) 
terminology is adopted here. 
 } 
 --- a spherically symmetric dust cosmology that is inhomogeneous in the 
radial direction.  If one accepts that the large scale universe is a 
collection of galaxies with negligible bulk rotation and interacting only 
through gravity, then the dust equation of state is valid, and the cosmic 
time evolution is pretty much determined along each worldline.  For 
reasons of simplicity we assume the observer is at the centre; thus we are 
assuming large scale isotropy about the milky way or some nearby point.  
Isotropy is relatively easy to test and is not a bad assumption on large 
angular scales.  It is also a natural simplification to make, since we are 
at the centre of the null cone we observe.  A more general approach would 
cloud the issue at hand, though it will be needed in the long run.  More 
importantly, the degree of deviation from isotropy is directly seen and is 
not mixed up with source evolution or cosmic evolution. 

 \section{The Cosmological Model} 

 We summarise the results of MHE, as background to the present 
considerations, and refer the reader to the papers cited there.  

 The 
 Lema\^{\i}tre-Tolman (LT) metric (Lema\^{\i}tre \cite{L33}, Tolman 
\cite{T34}, Bondi \cite{B47}) is 
 \begin{equation} 
   ds^2 = - dt^2 + \frac{(R')^2}{1 + 2E} \, dr^2 + R^2 \, d\Omega^2 \; , 
 \end{equation} 
 where $R = R(t, r)$, $E = E(r)$, $R' = \partial R/\partial r$, and 
$d\Omega^2 = d\theta^2 + \sin^2 \theta \, d\phi^2$.  The function $R$ is 
the areal radius, which obeys 
 \begin{equation} 
   \dot{R}^2 = \frac{2 M}{R} + 2 E 
 \end{equation} 
 and the arbitrary function $E$ determines both the local geometry and the 
local type of time evolution.  The arbitrary function $M = M(r)$ plays the 
role of the total gravitational mass interior to shells of coordinate 
radius $r$.  This has solutions in terms of the parameter $\eta$ 
 \begin{equation} 
   R = \frac{M}{\cal E} \, \phi_0(\eta) \; , ~~~~~~ 
      \xi(\eta) = \frac{({\cal E})^{3/2} (t - t_B)}{M} 
 \end{equation} 
 where 
 \begin{eqnarray}
 {\cal E}(r) 
 & = &
   \left\{ \begin{array}{l} 
   2 E(r) \; , \\ 
   1 \; , \\ 
   -2 E(r) \; , 
   \end{array} \right. 
 ~~~ \phi_0(\eta) = 
   \left\{ \begin{array}{l} 
   \cosh \eta - 1 \; , \\ 
   \eta^2 / 2 \; , \\ 
   1 - \cos \eta \; , 
   \end{array} \right. 
 \nonumber   \\
 ~~~ \xi(\eta) 
 & = &
   \left\{ \begin{array}{l} 
   \sinh \eta - \eta \; , \\ 
   \eta^3 / 6 \; , \\ 
   \eta - \sin \eta \; , 
   \end{array} \right. 
 ~~~ \mbox{when}~~~
   \frac{RE}{M} \, 
   \left\{ \begin{array}{l} 
   > 0 \; , \\ 
   = 0 \; , \\ 
   < 0 \; , \\ 
   \end{array} \right. 
 \end{eqnarray}
 and $t_B = t_B(r)$ is a third arbitrary function.  It gives the time of 
the big bang, $R = 0$, locally.  Although the initial singularity is 
spacelike everywhere%
 \footnote{ 
 except where there's a 
 shell-focussing singularity 
 } 
 the spacetime emerges from the initial singularity over a finite or 
possibly infinite time.  The density is 
 \begin{equation} 
   8 \pi \rho = \frac{2 M'}{R^2 R'} 
 \end{equation} 
 Putting the observer at the centre of symmetry, $r = 0$, the observer's 
past null cone is the solution of 
 \begin{equation} 
   dt = \frac{- R'}{\sqrt{1 + 2E}\;} \, dr 
 \end{equation} 
 that passes through $r = 0$ at $t = t_0$ 
 --- i.e. at the present.  MHE wrote this particular solution as $\hat{t} 
= \hat{t}(r)$, and a hat will henceforth indicate quantities evaluated on 
this null cone.  Since we really only need this one path, MHE used the 
freedom in the radial coordinate $r$ to specify 
 \begin{equation} 
   R' = 1 
 \end{equation} 
 on $t = \hat{t}(r)$ only.  This greatly simplifies the equations to be 
solved.  The path of the light cone is then 
 \begin{equation} 
   \hat{t}(r) = t_0 - r \; . 
 \end{equation} 
 MHE then went on to show how the 3 arbitrary functions of the LT model, 
$E(r)$, $M(r)$ and $t_B(r)$, could be determined from observations, if one 
knew the source evolution.  The observations needed were the source number 
counts against redshift $z$, and the diameter distance $R$ (or the related 
luminosity distance) against $z$.  The evolution functions needed were the 
mass per source, and the absolute diameter or luminosity, both against 
$z$.  Given only these observations, it was shown that both source 
evolution and inhomogeneity cause deviations from the expected FLRW 
observational relations%
 \footnote{ 
 Examples of strong deviations from the standard FLRW observational 
relations in models with realistic amounts of inhomogeneity were given in 
Mustapha Bassett Hellaby \& Ellis (\cite{MBHE98}) (MBHE).
 }, 
 and there is no way to distinguish the two. 

 We now generalise the MHE approach to observations in two or more  
different colours. 

 \section{Observing Model} 

 \subsection{Assumptions} 

 We assume the following: 
 \begin{itemize} 

 \item   The universe is isotropic about the earth to good approximation 
on the large scale. 

 \item   Small scale inhomogeneity has been eliminated by averaging cosmic 
observables over the whole sky, so that they are functions of $z$ only. 

 \item   There are $J$ types of visible sources, ~$1 \leq j \leq J$~, e.g. 
spirals, ellipticals, Abell clusters, field galaxies, etc.  The various 
types can be reliably distinguished.  The types have different intrinsic 
properties and {\em may} evolve differently 
 --- i.e. have different source evolution functions.  Mergers could 
correspond to the removal of certain types in favour of new types. 

 \item   There are $I$ different spectral frequencies or colour filters 
being used for observing, ~$1 \leq i \leq I$~, e.g. U, B \& V filters, or 
preferably a set of spectral line intensities. 

 \item   All wavelengths experience the same delay and the same 
cosmological redshift due to the geometry and evolution of the universe%
 \footnote{ 
 However redshifts of individual objects due to peculiar motions will 
differ, as noted below. 
 }%
 . 

 \end{itemize} 

 \subsection{Notation} 

 Whether the colour measurements are spectral line intensities or colour 
filter apparent magnitudes, we will call them ``colours''. 

 \subsubsection{Observables} 

 The following ``observables" are smooth functions fitted to data from 
direct observations, that have been corrected for selection effects, 
absorbtion and other effects mentioned below. 
 \begin{itemize} 

 \item   $z$ = measured redshift of each object. 

 \item   $n_j(z)$ = observed number density of source type $j$ in redshift 
space, per steradian per unit redshift interval, ~$1 \leq j \leq J$~. 

 \item   $\ell_{ji}(z)$ = measured apparent luminosity of source type $j$ 
in colour $i$ at redshift $z$, ~$1 \leq i \leq I$~. 

 \item   $\delta_{ji}(z)$ = measured angular diameter of source type $j$ 
in colour $i$ at redshift $z$. 

 \end{itemize} 

 \subsubsection{Theoretical Quantities} 

 The following are functions that must be supplied by a theory of source  
evolution. 

 \begin{itemize} 

 \item   $L_j(z)$ = absolute bolmetric luminosity of source type $j$, at 
the epoch corresponding to redshift $z$. 

 \item   $L_{ji}(z)$ = absolute luminosity of source type $j$, in colour 
$i$, at $z$. 

 \item   $D_{ji}(z)$ = proper diameter of source type $j$, in colour $i$, 
at $z$. 

 \item   $m_j(z)$ = total mass associated with (gravitationally bound to) 
source type $j$. 

 \item   $\nu(z)$ = total proper density in redshift space of matter not 
associated with a luminous source. 

 \end{itemize} 

 \subsection{Complications} 

 There are of course many sources of error in making and reducing the 
observations, which need careful attention.  However we will not go into 
them here, except to note the most important ones.  It is assumed the 
observational functions have already been corrected for these effects.  It 
is evident that measurements of spectral line intensities have fewer 
problems than luminosities through colour filters.  If a program of 
observations were being planned, line intensities are a natural extension 
to redshift measurements, and would be the natural choice.  However  
measurements of colour magnitudes are considerably easier to make, and 
more such data already exists. 

 \begin{itemize} 

 \item   The redshift we are interested in is the cosmological one, but 
the actual redshift of a source is the combination of cosmological and 
peculiar velocity contributions, causing considerable scatter in the 
observed $z$ values from a given distance.  Since the functions of $z$ we 
use here are 
 all-sky averages, this is not a problem.  However, in generalising to 
models with a measure of anisotropy in addition to radial inhomogeneity, a 
net `peculiar velocity' on the averaging scale being considered, should 
rather be viewed as an inhomogeneity in the cosmological expansion rate%
 \footnote{ 
 However, as was shown in MBHE, inhomogeneities near the maximum in 
$\hat{R}(z)$ can create loops in the $\hat{R}(z)$ graph, so that objects 
at different distances have the same $z$.  This is very like the ``finger 
of god'' effect.  Thus, if this were common, sources selected by $z$ may 
not all be at the same true distance, and there would be a blurring of 
source properties near the maximum, in addition to any observational 
uncertainties. 
 }%
 . 

 \item   The observed colours will be redshifted.  If spectral line 
intensities are used, then the emitted frequencies will be known, but if 
colour filters are used, different parts of the true spectrum are selected 
at each $z$ value, so correcting for this could be quite tricky, to say 
the least%
 \footnote{ 
 However, a well developed source evolution theory could make predicions 
for the relevant part of the spectrum according to $z$. 
 }%
 . 

 \item   Sources will suffer absorbtion and reddening.  This could be   
problematic as intergalactic absorbtion is not all that well known.  
Ideally wavelengths should be chosen to minimise absorbtion. 

 \item   Selection effects are crucial to real observations (Ellis, Perry 
\& Sievers \cite{EPS84}, Teerikorpi \cite{T97}, Totani and Yoshii 
\cite{TY00}).  Only a fraction $f$ of sources are actually detected, and 
$f$ depends on a source's apparent size and surface brightness.  Thus the 
observed number densities $n_j$ are related to the true number 
$\tilde{n}_j $ densities by 
 \begin{equation} 
 n_j = f(\ell_{ji}, \delta_{ji}, z) \, \tilde{n}_j 
 \end{equation} 
 For quantities calculated from the $n_j$, this is a serious problem.  But 
for those that depend on comparisons of {\em measured} $\ell_{ji}$ values 
 --- exactly what we need below 
 --- there is no real problem. 

 \item   Source evolution is naturally a function of time at the source, 
$\tau$, but here it is written as a function of $z$.  Although the age of 
the universe at each $z$ is an output function of the MHE algorithm, one 
needs the source evolution functions as initial inputs.  An iterative 
correction process could be developed, once our knowledge of source 
evolution is adequate to the task. 

 \item   It could be difficult to identify the same source type at 
different stages of evolution, especially at high $z$ values, and 
especially if galaxy mergers are very frequent.  For this reason it may be 
best to initially consider a single source type, and test for the bulk 
evolution properties of the cosmic luminous matter content. 

 \end{itemize} 

 \section{Comparison of Multicolour Observations} 

 Multicolour observations cannot be directly used as input to the LT model 
(or any other metric), which only requires $\hat{R}(z)$ and 
$\hat{\rho}(z)$ (plus the coordinate condition $\widehat{R'} = 1$) to 
fully determine the metric.  However for any given source, the ratio of 
apparent to absolute luminosity must be the same in all wavelengths.  In 
other words, the luminosity distances obtained in each of the colours or 
spectral lines must be the same. 

 \subsection{Distances} 

 The luminosity distance is 
 \begin{equation} 
   d_L = \sqrt{\frac{L}{4 \pi \ell}} 
 \end{equation} 
 Ideally, all estimates at any given $z$, for all source types and all 
colours, should agree, if we knew the $L_{ji}(z)$ completely: 
 \begin{equation} 
    \frac{L_{ji}(z)}{4 \pi \ell_{ji}(z)} = d_L^2(z) ~~~~~~ \forall ~ i, j 
 \end{equation} 
 This is a set of $I \times J$ equations for the $I \times J + 1$ unknown 
functions $L_{ji}(z)$ and $d_L(z)$.  Thus with sufficient observations we 
could determine the $L_{ji}(z)$ relative to each other, but none of them 
absolutely. 

 To get the function $\hat{R}(z)$ we take the average%
 \footnote{ 
 Equation (31) of MHE is incorrect. 
 } 
 \begin{equation} 
   \hat{R}(z) \, (1 + z) = \overline{d}_L ~,~~~~
   \overline{d}_L = \frac{ \sum_{j=1}^J \left( \frac{1}{I} \sum_{i=1}^I 
 \sqrt{\frac{L_{ji}}{4 \pi \ell_{ji}}} \; n_j \right)}{\sum_{j=1}^J n_j} 
    \label{lum-d} 
 \end{equation} 
 In reality, observations would be of different reliability, and a 
weighted average would be used. 

 We have similar relations for the diameter distance 
 \begin{equation} 
   d_D = \frac{D}{\delta} 
 \end{equation} 
 all of which should agree, if we knew the $D_{ji}(z)$ 
 \begin{equation} 
    \frac{D_{ji}(z)}{\delta_{ji}(z)} = d_D(z) ~~~~~~ \forall ~ i, j 
 \end{equation} 
 Again the average should give a good estimate of the LT areal radius 
 \begin{equation} 
   \hat{R}(z) = \overline{d}_D = 
   \frac{ \sum_{j=1}^J \left( \frac{1}{I} \sum_{i=1}^I 
 \frac{D_{ji}}{\delta_{ji}} \; n_j \right)}{\sum_{j=1}^J n_j} 
    \label{diam-d} 
 \end{equation} 
 and no absolute diameters can be determined. 

 If both types of distance measurements are available, then we have a 
combined requirement: 
 \begin{equation} 
   \frac{1}{(1 + z)^2} \frac{L_{ji}(z)}{4 \pi \ell_{ji}(z)} = \hat{R}^2(z) 
= 
   \left( \frac{D_{ji}(z)}{\delta_{ji}(z)} \right)^2 ~~~~~~ \forall ~ i, j 
    \label{both-d} 
 \end{equation} 
 leaving only one free function out of the set $\{ L_{ji}, D_{ji} \}$. 

  From here on we will focus on the luminosity distance, as the one that 
can be used to the 
  greatest depth. 

 \subsection{Densities} 

 If the number of sources of type $j$ observed between $z$ \& $z + dz$ in 
solid angle $d\Omega$ is 
 \begin{equation} 
   n_j \, d\Omega \, dz 
 \end{equation} 
 then the net mass is 
 \begin{equation} 
   n_j \, m_j \, d\Omega \, dz 
 \end{equation} 
 so the combined mass of all source types seen over the whole sky is 
 \begin{equation} 
   4 \pi \sum_{j=1}^J n_j \, m_j \, dz 
 \end{equation} 
 and if we allow for a distribution of 
 non-visible matter that isn't associated with any source, we get the 
total mass as 
 \begin{equation} 
   4 \pi \left( \sum_{j=1}^J n_j \, m_j + \nu \right) \, dz 
 \end{equation} 

  Now the local proper density in the LT model is 
 \begin{equation} 
   \rho = \rho(t,r) 
 \end{equation} 
 and its value on the null cone is 
 \begin{equation} 
   \hat{\rho} = \rho (\hat{t}(r), r) \; . 
 \end{equation} 
 The total mass between $r$ and $r + dr$ is 
 \begin{equation} 
   \hat{\rho} \, \widehat{d^3V} = \hat{\rho} \, \frac{4 \pi \hat{R}^2 
 \widehat{R'}}{\sqrt{1 + 2 E}\;} \, dr 
 \end{equation} 
 and hence 
 \begin{equation} 
   \hat{\rho} \, \frac{\hat{R}^2 \widehat{R'}}{\sqrt{1 + 2 E}\;} = 
   \left( \sum_{j=1}^J \, n_j \, m_j + \nu \right) \, \frac{dz}{dr} 
   \label{rho-n} 
 \end{equation} 

 \subsection{Fitting the Data with an LT Model} 

 Equations (\ref{rho-n}) and (\ref{lum-d}) or (\ref{diam-d}) or 
(\ref{both-d}) show how the LT functions relate to the observables and the 
source evolution functions.  Once the last two are known, the MHE 
procedure shows that, if we are given 
 \begin{equation} 
   \left( \sum_{j=1}^J \, n_j(z) \, m_j(z) + \nu(z) \right) 
   \mbox{~~~~~~and~~~~~~}   \overline{d}(z) 
 \end{equation} 
 then the 3 arbitrary functions that characterise the best fit LT model 
can be determined.  Thus {\em if} we knew the source evolution, we could 
determine the LT model that reproduces the observations.  In particular,  
we could determine how close the universe is to large scale homogeneity on 
the smoothing scale used. 

 \section{Number Counts} 

 The number counts only provide a single function of $z$ for each source 
type, and so do not help us to distinguish inhomogeneity from source 
evolution.  The advantage of the luminosity measurements is that each 
source is measured more than once in different frequencies. 

 \subsection{Colour-Weighted Number Counts} 

 Number counts of field galaxies with different colours have been used to 
try and demonstrate source evolution.  If such a method is undertaken, it 
is because the $\ell_{ji}$ are not known, and also because spectral line 
data is not available, only colour filter luminosities.  We here try to 
model this process in an admittedly rough and ready manner 
 --- the number counts obtained in different colours are modelled by 
weighting the total number count by a function of all the colour 
luminosities of that source type.  

 Let 
 \begin{itemize} 
 \item   $w_{ji}$ = weighting functions for each source type.  These may 
in general depend on all $I$ of the colour luminosities, e.g. $w_{23} = 
w_{23}(L_{21}, L_{22}, L_{23})$, 
 \item   $n_{ji}$ = number of source type $j$, counted in colour band $i$.  
Thus at each $z$ value: 
 \begin{equation} 
   n_{ji} = w_{ji}(L_{ji}) \, n_j 
 \end{equation} 
 \item   $c_i$ = 
 colour-weighted number count of all source types in colour band $i$ 
 --- i.e. the number density in redshift space $c_i(z)$ of all source 
types with colour $i$ is 
 \begin{equation} 
   c_i = \sum_{j=1}^J n_{ji} = \sum_{j=1}^J w_{ji}(L_{ji}) \, n_j 
 \end{equation} 
 \end{itemize} 
 The simplest weighting function is a step function with a 
 cut-off luminosity below which the source is not detected, 
 \begin{equation} 
   w_{ji} = \left\{ \begin{array}{ll} 
    1 & L_{ji} >    L_{ji}^0 \\ 
    0 & L_{ji} \leq L_{ji}^0 
   \end{array} \right. 
 \end{equation} 
 This means that each of the $c_i$ equals the total number count out to 
some $z$ value, and after that each source type disappears from the count 
fairly rapidly around some $z$ value.  The $z$ values for each type and 
each colour are probably not very different.  Again any variation could 
equally well be due to inhomogeneity in the population of source types as 
to source evolution.  Add to this the difficulty of correcting for the 
redshifting of frequencies detected in each colour band, and it is clear 
that this data does not assist in distinguishing source evolution from 
inhomogeneity.  Indeed, even if there were no large scale inhomogeneity, 
it is still difficult to get clear information about source evolution from 
this data%
 \footnote{ 
 For example, morphological studies suggest that the excess of faint blue 
galaxies is due to those with peculiar morphology.  But which of the more 
modern nearby galaxy types they evolve into, if any, or whether this is 
actually a manisfestation of cosmic inhomogeneity, is not known. 
 }%
 .  

 Now suppose the weighting function were proportional to the relative 
luminosity of that source type in that colour 
 \begin{equation} 
   w_{ji} = \frac{\ell_{ji}}{\overline{\ell}_j} ~~ , ~~~~~~~~ 
   \overline{\ell}_j = \frac{1}{I} \sum_{i=1}^I \ell_{ji} 
 \end{equation} 
 This is of course not realistic, but by presenting an excessively 
favourable case we highlight how unpromising the more realistic scenarios 
are.  For these weighting functions we would have 
 \begin{equation} 
   n_{ji} = \frac{\ell_{ji}}{\overline{\ell}_j} \, n_j 
          = \frac{L_{ji}}{\overline{L}_j} \, n_j ~~ , ~~~~~~~~ 
   \overline{L}_j = \frac{1}{I} \sum_{i=1}^I L_{ji} 
 \end{equation} 
 and so the number density in redshift space $c_i(z)$ of all source types 
with colour $i$ would be 
 \begin{equation} 
   c_i = \sum_{j=1}^J n_{ji} = 
         \sum_{j=1}^J \left( \frac{L_{ji}}{\overline{L}_j} \, n_j \right) 
 \end{equation} 
 where $\overline{\ell}_j$ and $\overline{L}_j$ are the mean luminosities.  
Assuming the $n_j$ are known, we have $I$ linear relations involving $I 
 \times J$ unknown coefficients.  This certainly doesn't help us solve for 
the evolution functions, though it does place mild constraints on them.  
The ideal situation is if there is only one source type, in which case we 
have $I$ equations for $I$ unknowns. 

 Any realistic weighting function would smear together these equations, 
making it virtually impossible to solve for the the coefficients with any 
certainty. 

 These results, even for the most ideal case, are {\it conditional on the 
number counts in each colour being tabulated against redshift $z$}.  But 
since they are {\it actually} summed over a large range of $z$, then the 
{\it constraints are lost}, as this effectively smooths out the evolution 
functions over time. 

 \section{Examples} 

 \subsection{1 Type of Source \& 3 Colours} 

 This is the case where we treat all galaxies as more or less the same, 
with roughly the same luminosity and colour evolution. 

We have $J = j = 1$, $I = 3$, $i = 1, 2, 3$, so all subscripts are colour 
subscripts.  The constraints are 
 \begin{equation} 
   \frac{L_1}{4 \pi \ell_1} = \frac{L_2}{4 \pi \ell_2} = 
   \frac{L_3}{4 \pi \ell_3} = \overline{d}_L^2(z) 
 \end{equation} 
 where all the $L$s and $\ell$s are functions of $z$, and 
 \begin{equation} 
   \overline{d}_L = \frac{1}{3} \left( \sqrt{\frac{L_1}{4 \pi \ell_1}}\; 
   + \sqrt{\frac{L_2}{4 \pi \ell_2}}\; + \sqrt{\frac{L_3}{4 \pi \ell_3}}\; 
   \right) 
 \end{equation} 
 This gives effectively 2 constraints on 3 evolution functions 
 --- e.g. $L_1/L_2$ and $L_2/L_3$. 

 \subsection{2 Types of Source \& 1 Colour} 

 In this case we can distinguish say 2 types of source, with different 
evolutions, but we only measure total luminosities. 

 We have $J = 2$, $j = 1,2$, $I = 1 = i$, so all indices are source type 
indices.  The constraints are 
 \begin{equation} 
   \frac{L_1}{4 \pi \ell_1} = \frac{L_2}{4 \pi \ell_2} = 
   \overline{d}_L^2(z) 
 \end{equation} 
 This clearly constrains the relative evolution of the source types. 

 \subsection{Several Types of Source and Several Colours} 

 If we need to distinguish 2 (or more) source types, and we make 
multicolour observations, then we get many more constraints, and we 
improve our chances of demonstrating source evolution.  If however the 
evolution of the source types is not very different, the extra effort 
would be unproductive.  Indeed, the more types we distinguish and the more 
colours we use, the smaller the sample size in each redshift interval.  
However, with the large scale sky surveys now operating or being 
developed, this may not be much of a limitation.  Nevertheless, 
identifying different source types requires resolved images or detailed 
spectroscopy for morphological classification, and involves considerable 
work. 

 \section{Conclusions} 

 \begin{itemize} 

 \item   The main result is that measurements of apparent luminosities in 
various colours $\{\ell_{ji}(z)\}$ 
 --- ideally a set of spectral line intensities 
 --- put strong constraints on the colour evolution functions 
$\{L_{ji}(z)\}$.  They determine all of the relative colour evolution 
functions $L_{ji}(z)/\overline{L}(z)$, but not $\overline{L}(z)$.  The 
same applies to the apparent diameters.  This allows source evolution 
theories to be tested against the observational data without any 
assumptions about homogeneity, or the cosmic equation of state.  

 \item  The essential point is, although luminosity distances and diameter 
distances are model dependent, affected by both inhomogeneity and equation 
of state, their ratios in different colours are not. 

 \item   The converse is not true 
 --- multicolour observations do not directly help to pin down the degree 
of inhomogeneity.  Once the {\it absolute} source evolution functions are 
reliably known from a well-confirmed galaxy evolution theory, a fit of the 
observational data would be possible. Otherwise, we would still need to 
know the true distances, by an independent method, such as supernova light 
curves, if they prove reliable, or gravitational microlensing. 

 \item   Colour band number counts $c_i(z)$ by themselves (without 
luminosity measurements) do not put much constraint on candidate evolution 
theories, even if they are known versus redshift $z$. 
 {\it But}, if they are summed over a range of $z$ values as is usual they 
tell us nothing. 

 \item   This method gives no direct information about the evolution 
quantities $m_j(z)$ and $\nu(z)$.  However they are more amenable to 
determination through measurements of orbital velocities in galaxies, 
galaxy interactions, gravitational lensing surveys, etc.  They will also 
be part of the evolution theories that are being tested by this method. 

 \item   The background model (LT) is not central to the ideas presented 
here.  The idea of comparing observational relations in different colours 
or frequencies to obtain clear evidence of source evolution will obviously 
apply in any model. 

 \item   As already emphasised, it is important {\em not} to test source 
evolution theories by fitting to a homogeneous FLRW model, because then 
the possibility of detecting large scale inhomogeneity 
 --- or of demonstrating homogeneity 
 --- is removed, and the theories are wrong if there really is 
inhomogeneity. 

 \item   The approach suggested here offers the possibility of testing 
evolution theories {\it independently} of whatever inhomogeneity may be 
present, which is a distinct advantage over other methods.  However, the 
determination of the inhomogeneity remains dependent on knowing the source 
evolution functions quite well. 

 \item   It is worth emphasising that obtaining 
 non-constant functions $L_{ji}(z)/\overline{L}(z)$ from multicolour 
observations does not itself prove the variation is due to time evolution
 --- it could equally well be due to spatial variation.  What it does do 
is provide data for testing theories of source evolution that is not 
contaminated by the gravitational effects of inhomogeneity in the 
intervening space.

 \item   The ideal observations envisaged 
 --- a redshift value and at least two spectral line intensities on a 
large number of galaxies 
 --- may be possible, with large scale observing programs along the lines 
of the SDSS, or with the new generation of telescopes: Keck, Subaru, 
Gemini, VLT.  The SDSS%
 \footnote{ 
 \tt http://www.sdss.org/ 
 } 
 is compiling an extensive and uniform database of galaxy data, for use in 
mapping the galaxy distribution, studying galaxy and quasar evolution and 
luminosity functions, improving the values of fundamental cosmological 
parameters, constraining dark matter distribution, and gravitational 
lensing studies.  When completed, it will have photometrically recorded a 
quarter of the sky in 5 colour bands between $3000$ \& $10\,000$ 
angstroms.  From this, $900\,000$ galaxies with mean $z \approx 0.1$ plus 
$100\,000$ luminous red galaxies with mean $z \approx 0.5$, and $100\,000$ 
quasars, will have been selected for spectroscopic imaging. 

 \item   We advocate an observational program, possibly using SDSS data, 
if suitable, aimed at extracting the source evolution functions, and 
ultimately establishing the degree of large scale inhomogeneity, through 
the approach suggested here, as a very worthwhile complement to other 
methods being pursued.  While there are plenty of practical problems 
involved, they are all familiar or being faced in projects currently under 
study. 

 \end{itemize} 

 \setcounter{secnumdepth}{0} 

 \begin{acknowledgements}

 I am grateful to South Africa's NRF for a research grant.  I thank George 
Ellis and the staff at the South African Astronomical Observatory, 
particularly Chris Koen, for helpful discussions, and I also thank 
Fernando de Felice for hospitality at the Dipartmento di Fisica of the 
Universit\`a degli Stud\^i di Padova while this paper was completed. 

 \end{acknowledgements}

 \end{document}